\documentclass[12pt]{article}

\global\arraycolsep=1pt

\usepackage{amsbsy,amssymb,latexsym,amsfonts, amsmath}
\usepackage{mathrsfs}
\usepackage{graphicx}

\usepackage{amsbsy,amssymb,latexsym,amsfonts,amsmath}
\usepackage{mathrsfs}
\usepackage{graphicx}
\usepackage{bm}
\usepackage{here}

\numberwithin{equation}{section}
\newcommand{\bel}[1]{\begin{equation}\label{#1}}                     
\newcommand{\bal}[1]{\begin{eqnarray}\label{#1}}                     
\newcommand{\bea}{\begin{eqnarray}}
\newcommand{\eea}{\end{eqnarray}}

\newcommand{\ex}{\mathrm{e}}

\renewcommand{\thefootnote}{\fnsymbol{footnote}}

\newcommand{\be}{\begin{equation}}
\newcommand{\ee}{\end{equation}}
\newcommand{\beq}{\begin{equation}}
\newcommand{\eeq}{\end{equation}}

\begin{document}
%
%
\begin{titlepage}
\begin{flushright}
\normalsize
~~~~
OCU-PHYS 448, v2\\
April, 2016 \\
\end{flushright}

\vspace{15pt}

\begin{center}
{\LARGE
Birth of String Theory}
\end{center}

\vspace{23pt}

\begin{center}
{ H. Itoyama$^{a, b}$\footnote{e-mail: itoyama@sci.osaka-cu.ac.jp} 
}\\
%
\vspace{18pt}
%

$^a$ \it Department of Mathematics and Physics, Graduate School of Science\\
Osaka City University\\
\vspace{5pt}

$^b$ \it Osaka City University Advanced Mathematical Institute (OCAMI)

\vspace{5pt}

3-3-138, Sugimoto, Sumiyoshi-ku, Osaka, 558-8585, Japan \\

\end{center}
%
\vspace{20pt}
\begin{center}
Abstract\\
\end{center}

This is a brief summary of an introductory lecture
for students and scholars in general given by
the author at Nambu Memorial Symposium which was held at
Osaka City University on September 29, 2015.
We review the invention of string theory by Professor Yoichiro Nambu
following the discovery of the Veneziano amplitude. We also
discuss Professor Nambu's proposal on string theory in the Schild gauge in 1976
which is related to the matrix model of Yang-Mills type.


\vfill

\setcounter{footnote}{0}
\renewcommand{\thefootnote}{\arabic{footnote}}

\end{titlepage}

\renewcommand{\thefootnote}{\arabic{footnote}}
\setcounter{footnote}{0}


\section{Introduction} \label{intro}

In 1949, Professor Yoichiro Nambu started his career at Osaka City University 
where this memorial symposium was held.  
The idea of this symposium was to deliver the remarkable scientific achievements 
and originality of Professor Nambu throughout his life as much as possible 
to younger generations and scholars in general, so that
some of what Nambu accomplished become more tangible.
On September 29, 2015, ten speakers contributed to the idea of the symposium.
The role of the author was to review the birth of string theory, which is a well-known
unpublished (not contradictory) work \cite{Namb1970D} of Nambu. For a pedagogical reason as well as
for the sake of presentation,  the developments based on the path integral method
before and after his work  were included. 
In the latter part of the talk, a less well-known proposal of Nambu
in the quantization and discretization of the string theory in the Schild gauge 
 in \cite{Schi1977C} was brought to the audience. This, in fact, testified for 
Nambu as a foreteller of modern physics. 
In order to set a current context of that proposal, 
 we include a general discussion of matrix models.

\section{Veneziano amplitude and Koba-Nielsen form} 
To begin, let us consider the $4$-point scattering of scalar mesons.
The Mandelstam variables are defined by 
\begin{align}
{s} &= (p_1+p_2) {\cdot} (p_1+p_2) \equiv (p_1+p_2)^2, \\
{t} &= (p_2+p_3) {\cdot} (p_2+p_3) \equiv (p_2+p_3)^2, 
\end{align}
where $p^{\mu}_I$ are the momenta and 
 the symbol $\cdot$ denotes the Lorentz invariant inner product in the Minkowski space.
\begin{figure}[H]
\centering
  \includegraphics[height=3cm]{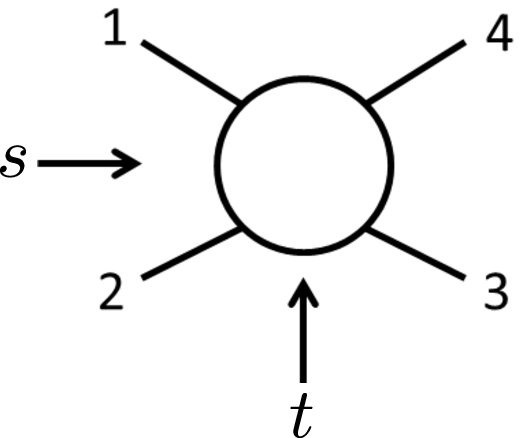}
\caption{ 
\label{scalar}
}
\end{figure}%
\noindent 
From experiments, the following facts were known: \\
$\bullet$ Contribution of many particles (poles) in the $t$ channel is evident. \\
$\bullet$ It suggests the structure of (mass)$^2$ proportional to spin. \\
$\bullet$ The behavior in the region $s \gg t$ is correlated with that in the region $s \ll t$.     

\begin{figure}[htb]
\centering
  \includegraphics[height=3cm]{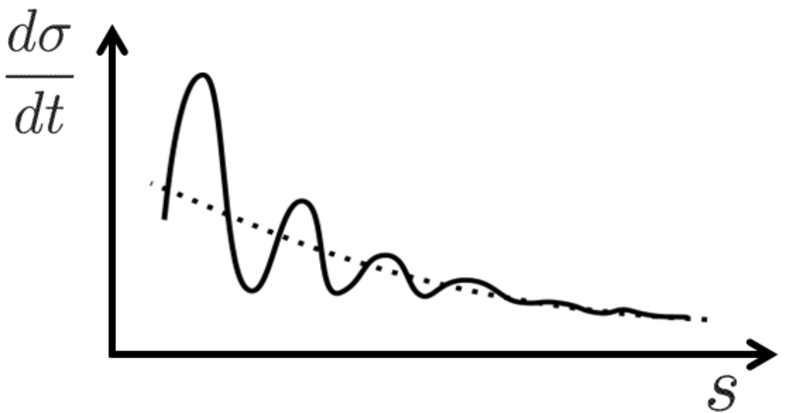}
\caption{ 
\label{graph}
}
\end{figure}%
\noindent 
 The idealization that one can take is that
there exists an infinite number of such particles (mass)$^2 \propto $ spin, 
 by setting $N_0 \to \infty$ in Fig. 3.
\begin{figure}[htb]
\centering
  \includegraphics[height=3.5cm]{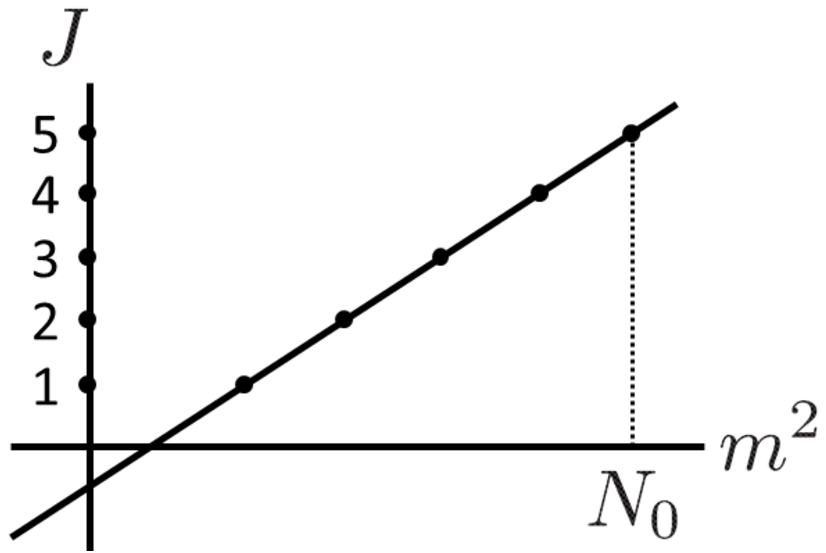}
\caption{ 
\label{graph2}
}
\end{figure}%

The amplitude was empirically known to satisfy the relation $f_{\rm exp}(s,t) = f_{\rm exp}(t,s)$, 
 which is  denoted by \lower2.3ex\hbox{\includegraphics[scale=0.4]{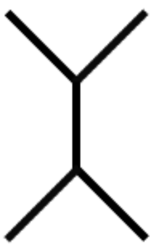}} 
$=$~\lower1.5ex\hbox{\includegraphics[scale=0.4]{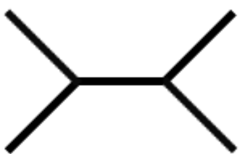}}. 
Therefore the behavior in the region where the first variable $\gg$ the second variable determines 
that in the region where the first variable $\ll$ the second variable. 
 The sum of exchange of the resonance in the $t$-channel 
 is equivalent to the sum of the resonance in the $s$-channel. 
  
Veneziano \cite{Veneziano} wrote down the amplitude satisfying these assumptions 
 and we will present the answer in a different way from his route.  We will also review the developments
where the solution was represented in such a way to be directly generalizable into 
$4 \to n$. 
\begin{figure}[H]
\centering
  \includegraphics[height=6cm]{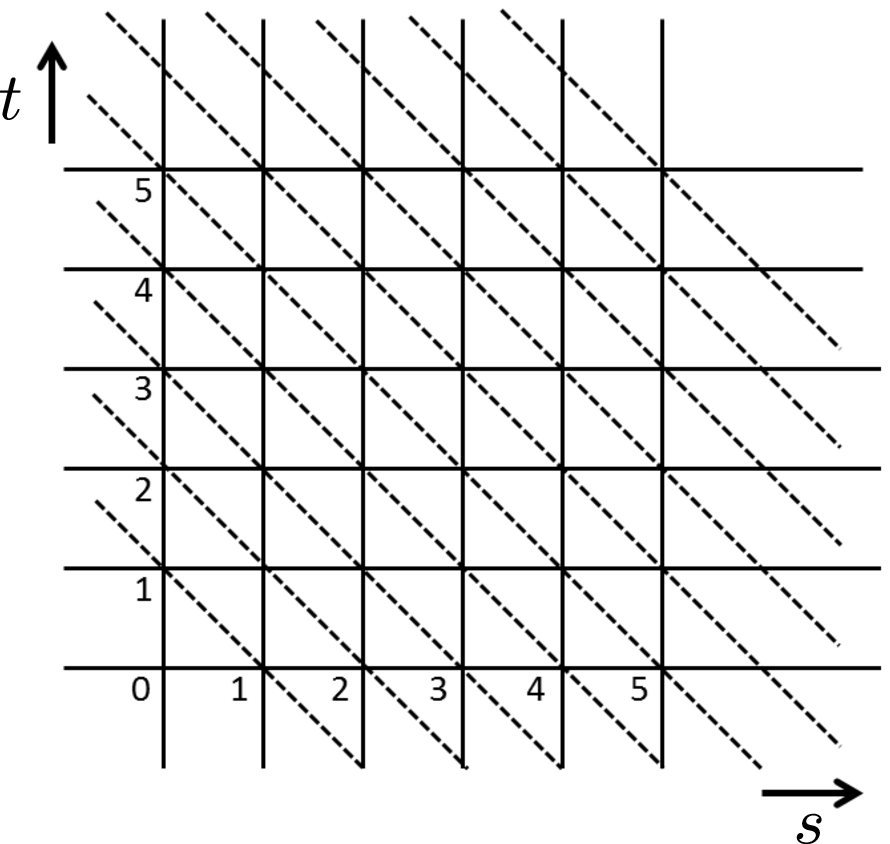}
\caption{ 
\label{fig}
}
\end{figure}%
For a while, let us keep $N_0$ in Fig. 3 finite and ignore the dimension, 
pretending that the slope of each trajectory is one. 
We need a simple pole at 
\begin{align}
\begin{cases}
 s=0, \dots, N_0, ~~~ t:~ \text{arbitrary} \\
 t=0, \dots, N_0, ~~~ s:~ \text{arbitrary}
\end{cases}
\end{align}
A naive multiplication of these pole factors, however, evidently leads us to 
 a problem of overcounting at these points where both $s$ and $t$ are integers. 
Multiplying the appropriate factors
indicated by the dotted lines in Fig 4 in order to remove this redundancy,
 we obtain
\begin{align}
&\frac{(-t-s)(1-t-s)\cdots(N_0-t-s) N_0!}
 {t(1-t)(2-t)\cdots(N_0-t) s(1-s) \cdots (N_0-s)} \cr 
&= \frac{N_0! (N_0-s-t)!}{(N_0-s)!(N_0-t)!} \frac{(-1-t)!(-1-s)!}{(-1-s-t)!}. 
\end{align}
Here, we have introduced the factor $N_0!$ to render the $N_0 \to \infty$ limit finite. 
Taking this limit,  we obtain
\footnote{The author came to know this derivation through B. Sakita. 
The derivation appears to have been known to some of workers at that time.}
\be
\xrightarrow[N_0 \to \infty]{}
{\frac{\Gamma(-t)\Gamma(-s)}{\Gamma(-s-t)}}. 
\ee
By the replacement $s \to \alpha(s) = \alpha_0 + \alpha' s$, we obtain 
\begin{align}
V_4 = \frac{\Gamma(-\alpha(s))\Gamma(-\alpha(t))}{ \Gamma(-\alpha(s)-\alpha(t))} 
 = B(-\alpha(s),-\alpha(t)) = \int_0^1 dx x^{-\alpha(s)-1} (1-x)^{-\alpha(t)-1}.
\end{align}

\begin{figure}[H]
\centering
  \includegraphics[height=2cm]{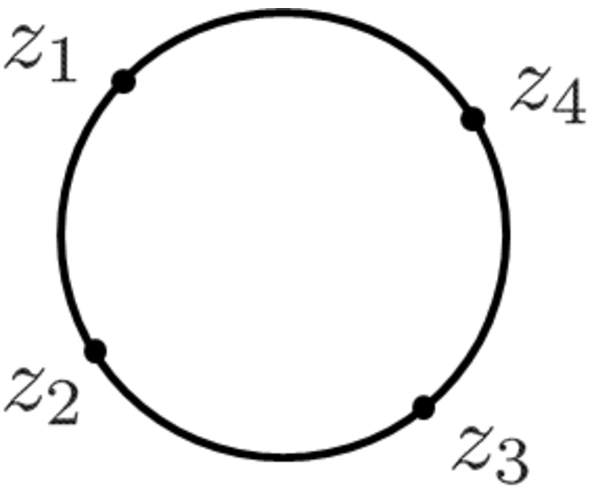}
\caption{ 
\label{circle}
}
\end{figure}%

The reduction of the number of integration variables from 4 $\to$ 1 variable $x$,
 is understood as  the M{\"o}bius invariance acting on the unit circle 
 in Fig. \ref{circle}. 
The cross ratio, defined by 
\begin{align}
{x = \frac{(z_1-z_2)(z_3-z_4)}{(z_2-z_4)(z_3-z_1)}}, 
\end{align}
and  we regard $z_2, z_3, z_4$ to be fixed by the M{\"o}bius group rotations.
 After some manipulations, we obtain 
\begin{align}\label{V4}{
V_4 = \int  \frac{\prod_{I=1}^4 dz_I/z_I}{d^3 F} 
 \prod_{I=1}^4 (z_{I+1}-z_I)^{\alpha_0-1} 
 \prod_{I > J} (z_I-z_J)^{2\alpha' k_I \cdot k_J}}, 
\end{align}
where 
\begin{align}{
d^3 F \equiv \frac{dz_2dz_3dz_4}{(z_2-z_3)(z_3-z_4)(z_4-z_2)}}. 
\end{align}
In this form, we can readily generalize the four point amplitude to 
the $n$ point  just by changing $4 \to n$ in the product. 
This is the Koba-Nielsen form \cite{KobaNielsen}.

Next, let us  rewrite the last expression \eqref{V4}, 
 using  the knowledge of two-dimensional
 electrostatics. Let us note that
\begin{align}\label{2.10}
(z_I - z_J)^{2 \alpha' k_I \cdot k_J} 
= \exp \biggl( 2 \alpha' k_I \cdot k_J \ln (z_I - z_J) \biggl),
\end{align}
 and that 
\begin{align}
N(z,\bar{z}; z', \bar{z}') \equiv \frac{1}{2\pi} \ln |z - {z}'||z-\bar{z}'^{-1}| 
\end{align}
is the Neumann function in the unit disk $D$.
We can rewrite  the exponent of the factor \eqref{2.10}
as
\begin{align}
\pi \alpha' \iint J(z) {\cdot}N(z,z') J(z'), ~~~~
 J^{\mu}(z,\bar{z}) = \sum_{I=1}^n k_I^{\mu} \delta^{(2)}(z-z_I),  
\end{align}
where 
\be
 z=\xi^1 + i \xi^2,  
\ee
and $z_I$'s are located on the boundary of the unit disk. 
It looks likes the (quantum)- oscillations of $d$ scalar fields 
 $X^{\mu}(z,\bar{z}),~ \mu= 0,1,\cdots, d-1$ on $D$ in two Euclidean dimensions 
 with such scalar fields path-integrated \cite{HSV1970F}. 
The action, which is a weight upon the path integrations, is identified as 
\begin{align}
{
 S = \frac{1}{2\pi \alpha'} \frac{1}{2} \int_{{D}} d^2 z \partial_{a} X \cdot \partial_{a} X}
~~~~~ a = \xi^1,~\xi^2.
\label{action:string}
\end{align}
Undesirable states, however, appear in the canonical quantization. 
An infinite number of constraints called Virasoro constraints \cite{Virasoro}
 must be imposed.

\section{Discovery of Nambu-Goto string} 
According to author's (certainly incomplete) search of references, several people other than
 the authors of \cite{HSV1970F} at that time  worked out the harmonic oscillator
 formalism to analyze the factorization and other properties of the Veneziano amplitude
and reached the quadratic form of the action \eqref{action:string}. 
 Most notably, the picture of rubber band was developed by Susskind 
 \cite{S1,S2,S3,NSK1971}.  
For contributions on these points from other people, including ones by unpublished reports, the author simply has no choice but to direct the readers to the references of the old
review article \cite{SCHERK} as well as those of the book \cite{GSWI}.

Nambu\cite{Namb1970D} and Goto\cite{Goto1971R} adopted 
 the area of the surface $\Sigma$ swept by a string
 in $d$ dimensional Minkowski spacetime as the action. 
The $d$ scalar fields introduced in eq. \eqref{action:string}  
 play the role of the embedding function: 
\begin{align}
X^{\mu}(z,\bar{z}) = X^{\mu} (\tau_M,\sigma), 
\end{align}
where $z = \xi^1 + i \xi^2 = e^{i(\sigma + \tau_M)}$. 
The action is given by 
\begin{align} \label{S_NG}
S_{NG} [X^{\mu}; \Sigma] = \frac{1}{2\pi\alpha'} 
\int_{\Sigma} d^2 \xi \sqrt{- \det \gamma_{ab}},
\end{align}
where $\gamma_{ab} \equiv \partial_a X \cdot \partial_b X$ is the induced metric. 
Since this action has  the reparametrization invariance, 
the following two constraints can be imposed:  
\begin{align}
\begin{array}{l}
 \displaystyle{T(\tau,\sigma) \equiv \frac{1}{2} (\dot{X} + X')^2} \approx 0
 , \\\\ \displaystyle{
 \bar{T}(\tau,\sigma) \equiv \frac{1}{2} (\dot{X} - X')^2} \approx 0,
\end{array}
\end{align}
where $\dot{X} \equiv \frac{\partial X}{d \tau_M}, ~
X' \equiv \frac{\partial X}{\partial \sigma}$. 
These are the Virasoro constraints. 
The action eq. \eqref{action:string} is reproduced, using these.

The next stage of the developments took place more than ten years later.
It came from our improved understanding of the quantization procedure based on
an auxiliary field (two dimensional metric) $g_{ab}(\xi)$.
The Nambu-Goto action is further rewritten 
in terms of intrinsic quantities of the two dimensional surface swept by a string:
\begin{align}
S_P[X^{\mu}, g_{ab}; \Sigma] &= \frac{1}{2\pi\alpha'} 
\frac{1}{2} \int_{\Sigma} d^2\xi \sqrt{-g} g^{ab} \partial_a X \cdot \partial_b X,  
\label{S_P}\\
g&= \det g_{ab}. 
\end{align}
This form is suited for the study of quantum anomaly 
\cite{Poly1981Q,Fuji1979P,Fuji1983P}.
Let us note that the action $S_P$ eq. \eqref{S_P} does not contain  derivatives of  $g_{ab}$. 
Equation of motion for $X^{\mu}$ is the two dimensional Laplace equation or  the wave equation, 
 while eq. of motion for $g_{ab}$ is the Virasoro constraints 
 or the energy-momentum tensor $= 0$  on the two-dimensional background metric. 
 The action $S_{NG}$ eq. \eqref{S_NG} is reproduced from $S_P$ by eliminating $g_{ab}$.

There is another way pursued on the quantization of the Nambu-Goto string
 in the so called Schild gauge:
\begin{align}\label{S_Schild}
``S_{\rm Schild}" \equiv \int d^2 \sigma \left( \alpha \sqrt{-g} 
 \frac{1}{4} \{X^{\mu},X^{\nu}\}_{\rm P. B.}^2 + \beta \sqrt{g} \right), 
\end{align}
where $\alpha$ and $\beta$ are constants and 
\begin{align}
\{X,Y\}_{\rm P.B.} \equiv \frac{1}{\sqrt{-g}} \epsilon^{ab} \partial_a X \partial_b Y
\end{align}
is the Poisson bracket. 
When eliminating $\sqrt{-g}$ by eq. of motion obtained from the variation $\delta{\sqrt{-g}}$, 
 the action $``S_{\rm Schild}"$ becomes that proportional to $S_{NG}$.

\section{Path integral quantization of string} 
Let us recall the bottom line of the path integral.
The transition amplitude for one-particle quantum mechanics is given by 
\begin{align}
\lim_{T \to \infty} \langle f| e^{-\frac{i}{\hbar} 
{\widehat{{H}}} T} | i \rangle 
= \int dx_f \int dx_i \psi^*_f(x_f) & \psi_i(x_i) 
{\sum_{\substack{\text{all paths (histories) under  } \\x(t=-\infty) 
 = x_i\\ x(t=+\infty) = x_f}}} e^{i {S[x]}/\hbar},
\end{align}
where $S$ is the action functional. 
This is essentially infinite dimensional multiple integrals. 

The path integral representation for the correlation function can be Wick rotated into
\begin{align}
\langle 0 ~ {\rm out} | \prod_{I=1}^n {\widehat{{O}}}_I(x_{t_I}) | 0~{\rm in} \rangle 
&=~~~~~\sum_{\substack{\text{all configuration fixed} \\ x(t_I) = x_{t_I} }} \prod_{I=1}^n O_I(x_{t_I}) e^{-S_E[x]}, 
\end{align}
which is the same as the sum over all configurations with Boltzmann weights in statistical
mechanics.

The M\"{o}bius invariant $n$-point scattering amplitude can be written as 
\begin{align}
\langle 0~{\rm out} | \prod_{I=1}^n \int \cdots \int \psi_I(z_I) \widehat{O}_I(z_I) | 0~ {\rm in} \rangle
&=\sum_{\substack{\text{sum over all surfaces} \\ \text{with weights} }}
\lower2.3ex\hbox{\includegraphics[scale=0.4]{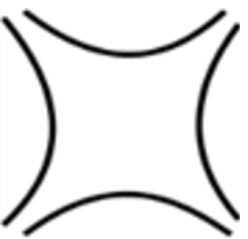}} .
\end{align}

Let us now turn to the path integral quantization of a string.
The path integrals over $X^{\mu}$ are essentially infinite-dimensional
 multiple integrals each of which is a gaussian.  
 The question arises on how to treat the path integrals for $g_{ab}$. 
Note that in computing $\sum_{{\text{configuration}}} \cdots e^{-S_E}$, 
the sum should be taken over the gauge inequivalent configurations alone.
The idea of the computation is that we first carry out the summation,
ignoring this double counting problem and then divide the answer by
``the number of multiplicities", namely, the volume of the gauge orbit. 
\begin{figure}[H]
\centering
  \includegraphics[height=3cm]{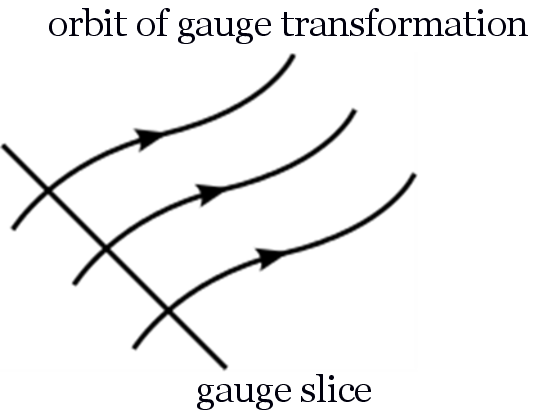}
\caption{ 
\label{gauge}
}
\end{figure}%
The general variation of $g_{\mu\nu}$ is given by
\be
 {\delta g_{ab} = (P_1 \delta v)_{ab} + \delta \rho g_{ab} +}
  { (\text{the variations not generated by the these two})}. 
  \label{variation:g}
\ee
The first one is the diffeomorphism with tracelessness condition imposed. 
The second term is the local Weyl transformation. 
These are the two local symmetries of the action \eqref{S_P}. 
The third term should be written as 
$\ker P_1^{\dagger}$ and corresponds to the degree of freedom called moduli 
of the surface deformation. 
Let us recall and indicate very briefly  how we introduce the notion of integration 
in finite dimensional Riemannian space with its metric given by 
\begin{align}
ds^2 = G_{AB} dY^AdY^b~~ \Rightarrow ~~ \mathcal{D} Y = \sqrt{G} \prod_A dY^A. 
\end{align}
We need to work out the infinite dimensional analog of this problem. 
Therefore, we have to estimate the Jacobian associated with the change of variables  
 \eqref{variation:g} \cite{Alva1983T,DH1988T}. 
 
Skipping the remaining procedure, let us give the master formula for string
perturbation theory for the case of a bosonic string. 
Here, we will consider the zero-point amplitude, 
namely, the partition function of a closed string only. 
The coupling constant of string theory is denoted by $\kappa$ and is introduced through
\begin{align}
e^{-\frac{1}{4\pi} \int d^2 \xi \sqrt{-g} R^{(2)} \Phi} \equiv \kappa^{- \chi(h)}, 
\end{align}
where $\Phi$ is the condensate of dilaton.
The order of string perturbation theory is determined 
by the Euler number $\chi(h,b,c)$ of the surface
($h$, $b$ and $c$ are the number of holes, boundaries and 
 cross caps, respectively) and 
string perturbation theory is, therefore, the genus expansion. 
The formula for the partition function reads  
\begin{align}
Z = (\text{$d$-dim. vol.}) \sum_{h=0}^{\infty} \kappa^h \int_{{({\rm moduli})_h}} 
 d(WP)_{\hat{g}} \frac{(\det' \hat{P}_1^{\dag} 
 \hat{P}_1)^{\frac{1}{2}}}{{\rm vol}({\rm Ker} P_1)} \left(
\frac{2\pi}{\int d^2 \xi \sqrt{\hat{g}}} \det' \Delta_{\hat{g}}
\right)^{-\frac{d}{2}}   
\int \mathcal{D} \phi e^{-(26-d) S_L(\phi,\hat{g})}. 
\end{align} 
For the detail of the notation, we refer the reader to
\cite{Alva1983T, DH1988T}.
The critical dimension $d=26$ is selected
by demanding the decoupling of the Liouville degrees of freedom.

In the case of superstring perturbation theory, 
we first need to introduce fermions on the world-sheet 
\cite{NS1971F,Ramo1971D,IK1973Q} 
but we need to work out a lot more to construct 
the perturbative series. 
The critical dimension is $d=10$.
Maximal spacetime supersymmetry is accomplished by the GSO projection \cite{GSO1977S} 
 and the generalized GSO projection \cite{KLT1986C,KLT1987C,LLS1987}  
 or orbifolding \cite{DHVW1985S} 
must be made in order to construct more realistic models having non-maximal 
supersymmetry. 
Eventually, spacetime supersymmetry must be broken and currently 
there is a revived interest (see, for instance, \cite{ADM1502T}) 
in the old work \cite{SS1979H,DH1986S,IT1987S}.
Turning to the more mathematical aspects, the construction of super Riemann surfaces
(see, for instance, \cite{Rabi1987T,ARS1988A,RSV1988G,BM1987O,IM1987M})   
 as well as that of the super moduli has been major unfinished parts. 
See, for instance, \cite{DW1304S} for recent progress.

\section{Matrix model of Yang-Mills Type and Nambu's Proposal}
  Matrix models are defined by
\begin{align}
Z(\text{parameter}) = \iint \cdots \int \prod_I \prod_{i,j} 
 d M_{ij}^{(I)} e^{-S(M^{(I)};\text{parameter})}. 
\end{align}
These are just finite dimensional multiple integrals. There are two types:
1) the one matrix model and its extension to a chain of matrices. 
2) the Yang-Mills type. 

The Haar measure of an $N \times N$ Hermitian matrix $M$ is given by
 \begin{align}
dM = \prod_{i=1}^N d\lambda_i \prod_{i>j} 
 d {\rm Re}\Omega_{ij} d {\rm Im} \Omega_{ij} 2^{\frac{N-1}{2}} 
 \prod_{i>j}(\lambda_i-\lambda_j)^2, 
\end{align}
where 
$M = U^{\dag} \Lambda U$ and $d\Omega = dUU^{\dag}$. 
The factor $\prod_{i>j}(\lambda_i-\lambda_j)^2$  works as a repulsive force 
 between the eigenvalues in providing the effective action of the model given. 
 The eigenvalue distribution is expected to become continuous in the limit $N \to \infty$ and
it can be regarded as a system of complex planes cut and glued, namely, the Riemann
surface.  

In the case of the one-matrix model, the simplest model of type 1), 
 equation of motion for the correlation functions 
 (the Schwinger-Dyson equation) takes the form of  the Virasoro constraints 
 \cite{Davi1990L,IM1991N,MM1990O,AJM1990M}.   
For more general chain models, they typically obey $W_n$ type constraints
 \cite{MMM1991G}. 
The models of type 1) reduce to eigenvalue models as the angular
integrations simply factor out. 

There are several physical contexts where the models of type 1) are relevant:
\begin{enumerate}

\item The string theory where the Liouville 
factor $g^a_a = 2\ex^{\phi}$ cannot be factorized 
 \cite{Davi1990L,DK1989C} (string theory in non-critical dimensions). 
Furthermore, in the case of $d \leq 1$, one can sum the perturbative series 
to treat some non-perturbative effect  \cite{BK1990E,DS1990S,GM1990N}.

\item Choosing a multi-log potential and an appropriate integration region 
(Selberg-type matrix model), and introducing the parameter $\beta$ to modify 
the exponent of the measure factor ($\beta$-ensemble),
the instanton sum has been generated (AGT relation \cite{AGT0906L})
\cite{IO1003M,DV0909T}.

\item Generation of the effective (super-)potential which 
 describes the gluino condensation  \cite{DV0206M,IM40211T}. 

\item Actions of $d=3$ and $d=4$ supersymmetric gauge theories often reduce to matrix
integrals (localization). (See, for example, \cite{Pest0712L}). This has led to
the study of instanton gas of various kind. (See, for example, \cite{HMO1207E}).  

\end{enumerate}

With regard to the matrix models of type 2), Professor Nambu made
a remarkable proposal already in 1977 in \cite{Namb1977S}
in string theory in the Schild gauge. In fact, equation of motion for $X^{\mu}$
obtained from  $``S_{\rm Schild}"$ in eq. \eqref{S_Schild} is  
\begin{align}
\left\{X_{\mu}, \{X^{\mu} , X_{\nu}\}_{\rm P.B.} \right\}_{\rm P.B.} = 0.
\end{align}
Let us quote Nambu's proposal:
\begin{quote}{
\textit{``An interesting possibility that suggests itself is to take 
the Poisson bracket notation in Eq. (35) seriously, and go to its 
``quantum mechanics version", by regarding the internal coordinates
  $\tau$ and $\sigma$ as non-commuting operators. 
 It is totally unclear what this means, but we try it nevertheless. \dots"
}}\cite{Namb1977S} 
\end{quote}

When the Poisson bracket $\{~,~\}_{\rm P.B.}$ and $X^{\mu}$ 
 are replaced by $-i [~,~]$ and by the covariant derivative   
 $D^{\mu} = \partial^{\mu} - igA^{\mu}$, respectively, we in fact obtain
an equation of motion for Yang-Mills fields 
\begin{equation}
[D_{\mu}, [ D^{\mu},D^{\nu}]] = 0.
\end{equation}

The IIB matrix model  \cite{IKKT9612A} was proposed in order to provide a complete treatment 
 of superstrings that includes non-perturbative effects. 
 The action is given by 
\begin{align} {
 S_{\rm IKKT} = - \frac{1}{g^2} {\rm Tr} \left(
 \frac{1}{4} [A_{\mu},A_{\nu}][A^{\mu},A^{\nu}] + 
 \frac{1}{2} \overline{\psi} \Gamma^{\mu} [A_{\mu}, \psi] \right)},  
 \label{IKKT}
\end{align}
where $A^{\mu}~ (\mu=0,1,\cdots,9)$ are ten Hermitian matrices and 
 $\psi_{\alpha}$ is a 16-component Majorana-Weyl spinor that takes values in 
 $N\times N$ Hermitian matrix. We have denoted by
  $\Gamma^{\mu}$ 10-dimensional gamma matrices. 
The action \eqref{IKKT} is obtained 
 from the Green-Schwarz action ($\theta_2 \to i\theta_2$) in the Schild-type gauge 
 by the following replacements,  
\begin{align}
\{~~,~~\} ~~&\to~~ -i[~~,~~], \\
\frac{1}{2\pi} \int d^2\sigma \sqrt{-g} ~~&\to~~ {\rm Tr}. 
\end{align}
The bosonic part takes the same form as that proposed by Nambu. 
The many-body problem of strings can be treated by
integrating out the off-diagonal blocks after dividing each of the
original matrices into blocks. 
It appears, however, that there is no evidence that the gauge volume 
of the local Weyl symmetry is factored out, which is requisite
for (perturbative) string theory in the critical dimension that ensures masslessness
of graviton in flat spacetime.

\section*{Acknowledgement}
The author (H.I.) was neither a postdoc nor a graduate student of Professor Yoichiro
Nambu's, but had a rare fortune of being relatively near at FNAL-Chicago, 
 Osaka University and Osaka City University in the last thirty years, 
 and could know some of his thoughts and inspirations in depth. 
 The author expresses his gratitude to Professor Nambu for providing him insights 
 as a scientist and tenderness as a human in various occasions. 


\end{document}